# Balancing operator's risk averseness in model predictive control for real-time reservoir flood control

Ja-Ho Koo [a,b,*], Edo Abraham [b], Andreja Jonoski [a] and Dimitri P. Solomatine [a,b,c]

[a] Department of Hydroinformatics and Socio-Technical Innovation, IHE Delft, Westvest 7, Delft, AX 2611, The Netherlands
[b] Department of Water Management, Faculty of Civil Engineering and Geosciences, Delft University of Technology, Delft, CN 2628, The Netherlands
[c] Department of River Basins Hydrology, Water Problems Institute of RAS, Gubkina 3, Moscow 119333, Russia
*Corresponding author. E-mail: j.h.koo@tudelft.nl

J-HK, 0000-0001-7100-8518; EA, 0000-0003-0989-5456; AJ, 0000-0002-0183-4168; DPS, 0000-0003-2031-9871

## ABSTRACT

Model predictive control (MPC) is an optimal control strategy suited for flood control of water resources infrastructure. Despite many studies on reservoir flood control and their theoretical contribution, optimisation methodologies have not been widely applied in real-time operation due to disparities between research assumptions and practical requirements. To address this gap, we include practical objectives, such as minimising the magnitude and frequency of changes in the existing outflow schedule. Incorporating these objectives transforms the problem into a multi-objective nonlinear optimisation problem that is difficult to solve in real time. Additionally, it is reasonable to assume that the weights and some parameters, considered the operators' preferences, vary depending on the system state. To overcome these limitations, we propose a framework that converts the original intractable problem into parameterised linear MPC problems with dynamic optimisation of weights and parameters. This is done by introducing a model-based learning concept. We refer to this framework as Parameterised Dynamic MPC (PD-MPC). The effectiveness of this framework is demonstrated through a numerical experiment for the Daecheong multipurpose reservoir in South Korea. We find that PD-MPC outperforms standard MPC-based designs without a dynamic optimisation process for the objective weights and model parameter. Moreover, we demonstrate that the weights and parameters vary with changing hydrological conditions.

Key words: dynamic preferences, flood control, large reservoirs, model predictive control, operators' preferences, practical application

## HIGHLIGHTS

- Detailed practical objectives for real-time reservoir flood control are explored and suggested.
- We propose the use of dynamic weights and parameters of objectives when changing operators' preferences are assumed.
- We propose a PD-MPC framework as a parameterised linear MPC with dynamic optimisation of weights/parameters.
- Numerical experiment reveals PD-MPC outperformed MPC with fixed weights/parameters.

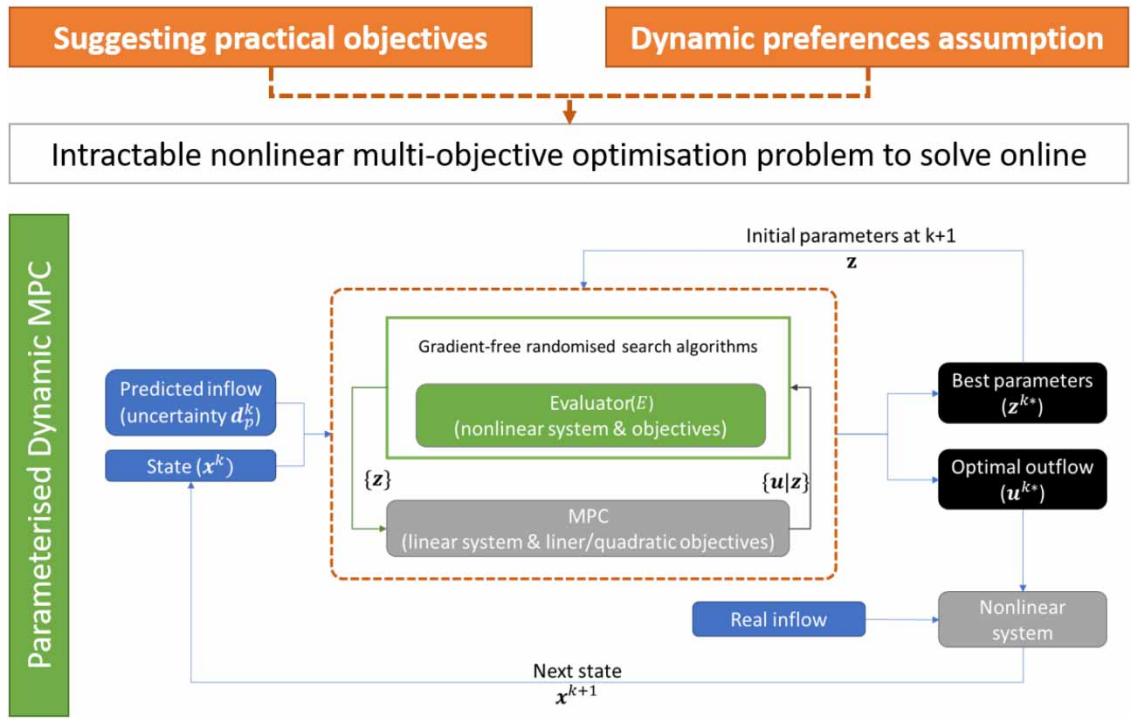

DP      dynamic programming
FWL     flood water level
GA      genetic algorithm
LP      linear programming
LWL     low water level
MAVE    maximum allowed value estimate
MPC     model predictive control
NHWL    normal high water level
PD-MPC  parameterised dynamic model predictive control
RWL     reservoir water level

## 1. INTRODUCTION

Many multipurpose reservoirs are designed and constructed with flood control as an important objective. They play a crucial role in mitigating flood risks downstream by retaining a significant portion of inflows in a given period. Additionally, the efficient operation of existing reservoirs is becoming increasingly important since costs and growing environmental concerns in society are making it challenging to build new reservoirs (Scudder 2012). At the same time, climate change may lead to more severe and unpredictable flood events (Havens *et al.* 2016), which are difficult for operators to manage efficiently using conventional simulation-based methods (Watts *et al.* 2011).

Decisions regarding flood control operations for such reservoirs significantly influence basin flood conditions. Due to the importance of reservoir flood control, achieving *optimal* outflows of reservoirs has long been a focus (Giuliani *et al.* 2021; Jain *et al.* 2023). Various optimisation methodologies, including linear programming (LP), dynamic programming (DP), and their variants, have been widely applied due to their capacity to guarantee optimal solutions (Labadie 2004) for this type of problem. Another popular class of algorithms in the literature on reservoir optimisation is randomised search techniques, mainly evolutionary algorithms and the genetic algorithm (GA). These have also gained traction, particularly for addressing complex problems that are not analytically expressed and less tractable to pose as mathematical optimisation problems (Ahmad *et al.* 2014). Some relatively recent literature has also focused on the approaches associated with control theory, primarily model predictive control (MPC), with different optimisation algorithms employed to solve the resulting optimal control problem (Breckpot *et al.* 2013; Delgoda *et al.* 2013; Castelletti *et al.* 2023).

Reservoir operators use only a limited horizon of rainfall/inflow forecasts, which are often uncertain. These forecasts are typically shorter than the whole length of a flood event (Breckpot *et al.* 2013). Operators must make decisions based on uncertain short-term forecasts. These decisions are not only about one outflow for the current time but also outflows for some time horizon, i.e., an outflow schedule. However, operators implement only the first (sometimes a few) outflows in the outflow schedule because, in general, forecasts are updated regularly, and new forecasts are considered less uncertain. Therefore, decisions should be updated repeatedly to reflect updated forecasts and hydrological conditions. In this sense, the receding horizon MPC concept, in which only the first control input is implemented and the optimisation processes are repeated at each time step iteratively (Van Overloop 2006), coincides with the reservoir flood control.

While optimisation-based control approaches including MPC have made substantial contributions in many applications (Schwenzer *et al.* 2021), their practical application within reservoir flood control remains somewhat limited. This discrepancy can be mainly attributed to the disparities between research assumptions and the pragmatic necessities of real-time reservoir flood control.

The first disparity concerns objectives. Many researchers have introduced objectives for optimal reservoir operation (Ko *et al.* 1992; Reddy & Kumar 2006; Malekmohammadi *et al.* 2011; Lin & Rutten 2016; Tang *et al.* 2019), such as minimising outflows via spillway gates, maximising hydropower generation, and maintaining reservoir water level (RWL) for water supply. However, during a flood event, most of these objectives may not be a priority.

What operators want in practice for flood control, i.e., operators' objectives for short-term planning, is not generally specified in the regulations and guidelines but is considered ambiguous 'experience' or 'expert knowledge'. Here, we call these 'practical objectives'. Note the distinction from mathematical objective functions, where mathematical equations, including system models and constraints, are used to calculate values of control objectives. Researchers seem to have paid insufficient attention to how to define and adopt these practical objectives in the optimisation process (Ritter *et al.* 2020; Teegavarapu & Simonovic 2001). A recent survey conducted on water supply companies in England and Wales by Pianosi *et al.* (2020) has shown that factors, i.e., objectives and constraints, in decision-making were too complex to be included in an optimisation process. Hence, operators generally hesitated to adopt optimisation tools. Moreover, even if we successfully formulate practical objectives but with a number of highly nonlinear mathematical formulas, the problem then becomes intractable by the MPC approach (Allgower *et al.* 2004; Berberich *et al.* 2022). This is often a reason why many researchers apply only a limited number of objectives or linear and/or quadratic objectives, which are tractable even with several objectives, for a reservoir flood control problem (Breckpot *et al.* 2013; Qi *et al.* 2017; Uysal *et al.* 2018b) and so are deemed feasible to solve.

In addition, the preferences imply the selecting criteria of operators among control alternatives, which are calculated from different weight sets and/or parameters. Therefore, selected weights and parameters in the multi-objective optimisation problem can be referred to as the *preference* (Wang *et al.* 2017). In many studies focusing on reservoir flood control and applying the receding horizon MPC approach, weighted multi-objective optimisation-based methods have been adopted (Wang *et al.* 2013; Hu *et al.* 2014). Hence, it is necessary to produce a Pareto set, which is a vector of non-dominant control inputs, using a weighted-sum method or apply one fixed weight representing the operators' preference, i.e., allocating high weights to the objective to which operators pay more attention. However, producing a Pareto set at every time step is computationally expensive (Peitz & Dellnitz 2018). Additionally, we cannot be sure that the relative importance of objectives remains constant during a flood event.

To fill these research gaps, this article presents an MPC framework to generate an optimal flood control decision, which, in our opinion, would be practical and more acceptable to operators. The main idea is explicitly defining the practical objectives for reservoir flood control based on the first author's operational experience. We assume dynamic preferences, which we then formulate using parameterised linear MPC and dynamic optimisation of weights and parameters of objectives to efficiently handle an otherwise intractable multi-objective nonlinear optimisation problem. This methodology is tested for historical flood events with different features, such as the number of inflow peaks and event lengths. Such verification is important because outflows and water level at the end of the first/second peak hugely influence the outflows of the next time steps.

The manuscript is organised as follows. In Section 2, we present the objective functions for practical flood control and propose a framework to efficiently incorporate linear and nonlinear objectives under the assumption of dynamic preference. Section 3 describes the detailed MPC problem and a numerical experiment. The contribution and limitation of this research and the result of the numerical experiment are then presented in Section 4, followed by conclusions.

## 2. METHOD

### 2.1. Model predictive control

Model predictive control (MPC) is an optimal control framework that uses a system model to predict and evaluate the system's behaviour over a finite time horizon. The control input, which is the optimal outflow schedule in reservoir flood control, is calculated to optimise objectives and satisfy constraints based on the current state and predicted behaviour. The main characteristic of the MPC approach is the combination of prediction and optimisation. Only the first outflow in the schedule is implemented, and the prediction and optimisation processes are repeated at each time step. This iterative process is generally referred to as receding horizon control (Van Overloop 2006). In some applications, such an approach is also called a rolling horizon method (Wang *et al.* 2014).

Unlike randomised search algorithms such as GA, MPC can explicitly consider constraints. Another significant advantage of MPC is its robustness to disturbances, such as uncertain inflow forecasts and system uncertainty. This robustness stems from MPC's ability to re-estimate predictions and recalculate optimal control inputs at each time step based on updated hydrological states, including revised inflow forecasts (Schwenzer *et al.* 2021).

### 2.2. Objective functions for practical reservoir flood control

The flood control of large reservoirs requires multiple practical considerations; one well-known conflicting objective is the need to reserve enough water to supply to contractors at the end of a flood event. In this section, we propose additional important objectives and motivate their necessity based on the receding horizon MPC concept.

First, using the full capacity of control facilities such as spillway gates is not reasonable for large reservoirs. Instead, operators tend to desire less outflow generally. Second, it is preferable to limit the frequency of operations of spillway gates to prevent wear and malfunction. Furthermore, immediate and frequent changes in outflow schedules are not preferred because they can hinder the predictability of the flood situation of the downstream area for other flood control agencies. It is worth noting that even though we are focusing on flood-relevant objectives, other objectives for reservoir operation, such as maintaining minimum flow requirements or ensuring adequate flow for fish migration and downstream habitats, can be formulated similarly.

In defining the practical objectives and the subsequent methodology, we introduce a control input vector, $\mathbf{u}^k$, at time step $k$, defined as $\mathbf{u}^k = \{O_{total,k}^k, \ldots, O_{total,k+H-1}^k, O_{spill,k}^k, \ldots, O_{spill,k+H-1}^k\} \in \mathbb{R}^{2H}$ where $O_{total,k+H-1}^k$ and $O_{spill,k+H-1}^k$ are the total outflow and spillway outflow decided at time step $k$ for time $k+H-1$, respectively. $H$ is the length of the prediction horizon. We define an augmented state vector that consists of the RWL, predicted inflow, and the outflows decided at the previous time step $k-1$ as $\mathbf{x}_k = \{S_k, I_k, \ldots, I_{k+H-1}, O_{total,k}^{k-1}, \ldots, O_{total,k+H-1}^{k-1}\} \in \mathbb{R}^{1+2H}$, where $S_k$ and $I_k$ are the storage volume and predicted inflow variables for time step $k$, respectively. To avoid ambiguity, we define the term 'time step $k$' as the $k$th iteration of MPC and 'time $t$' as the exact hour at which a control input is supposed to be implemented. The detailed objectives for practical reservoir flood control are defined below.

#### 2.2.1. Minimising the peak and total outflow via spillway gates

Operators typically prefer using turbines over spillways due to the two objectives of reducing the risk of flood damage in the downstream area and maximising earnings from hydropower generation, respectively. Generally, these two objectives are not in conflict but rather complementary. For example, to reduce the peak outflow, it is essential to distribute outflows over time, which consequently leads to maximising turbine outflows and minimising the total volume of spillway outflows. Therefore, these objectives could be applied both together or separately; at least one of them appears to be necessary in any case. The objectives can be formulated as follows:

$$\tilde{J}1 := \min_{\mathbf{u}^k} \max_{\forall t \in \{k, \dots, k+H-1\}} O^k_{spill,t}, \text{ and} \tag{1}$$

$$\tilde{J}2 := \min_{\mathbf{u}^k} \sum_{t=k}^{k+H-1} O^k_{spill,t}. \tag{2}$$

The first equation minimises peak outflow in a prediction horizon (from $k$ to $k + H - 1$ at time step $k$), which is $\max_{\forall t \in \{k, \dots, k+H-1\}} O^k_{spill,t}$, where $O^k_{spill,t}$ means a spillway outflow at time $t$ decided at time step $k$. The second one minimises the sum of spillway outflows, $\sum O^k_{spill,t}$, during a prediction horizon.

### 2.2.2. Minimising step-wise outflow changes in the prediction horizon

For large reservoirs, operators generally hesitate to change outflows drastically. Rather, it is preferred to implement smooth and deliberate adjustments, avoiding significant and sudden changes. This can be achieved by minimising cumulative step-wise changes over a given receding horizon as:

$$\tilde{J}3 := \min_{\mathbf{u}^k} \sum_{t=k}^{k+H-2} |O^k_{total,t+1} - O^k_{total,t}| \times w_t, \tag{3}$$

where the absolute value notation could be replaced with a quadratic of the step-wise changes. However, the use of absolute value is preferred to also penalise small changes in outflows. To avoid excessive penalisation for every change, we can assign greater weight to changes occurring at shorter time intervals than the changes in outflows at longer intervals by introducing weights $w_t$ in this objective.

It is noteworthy that $\tilde{J}3$ and $\tilde{J}2$ complement each other to generate an outflow sequence with less variability. When outflows need to increase, e.g., when RWL is expected to rise significantly due to substantial inflow, $\tilde{J}2$ leads to dropping the last outflow in the outflow schedule to minimise the total spillway discharge over the prediction horizon. This occurs because, in the discretised reservoir model, which will be detailed in Section 3.2.1, the last outflow does not affect RWLs within the same prediction horizon, as outflows and inflows at a given time, e.g., $t$, only influence the reservoir's hydrological condition at the subsequent time, $t + 1$. $\tilde{J}3$ can prevent this undesirable schedule formation. Conversely, when outflows need to decrease, e.g., when inflows are insufficient to maintain RWLs for water supply despite the earlier prediction of substantial inflows, $\tilde{J}3$ may impede the reduction of the final outflow, but $\tilde{J}2$ can enforce this reduction. This complementarity may be deemed less crucial when the prediction horizon is long, given the diminishing significance of outflow changes distant from the current time step. However, in scenarios with a short prediction horizon, such as the 6-h case in our numerical experiment in Section 3, these objectives can be essential for making the optimal outflow practically applicable.

### 2.2.3. Minimising changes in outflows calculated at consecutive time steps

In receding horizon MPC, the optimal control inputs are generated at each time step for the prediction horizon. In many MPC applications, previous optimal control inputs are not taken into account at the next time step calculations. However, because the optimal control inputs decided at the previous time step are an *outflow schedule*, it is also crucial to minimise the changes in each control input over each prediction horizon to prevent sudden alterations in outflow schedules. This objective can be formulated as:

$$\tilde{J}4 := \min_{\mathbf{u}^k} \sum_{t=k}^{k+H-2} |O^k_{total,t} - O^{k-1}_{total,t}| \times w_t, \tag{4}$$

where $w_t$ are weights. The previous objective, Equation (3), which aims to minimise the step-wise changes within a prediction horizon, serves a similar purpose to Equation (4) when outflows do not change much. To avoid excessive penalisation for every change, the weight $w_t$ is also applied for the same reason as in Equation (3).

## 2.2.4. Minimising the RWL exceedance outside of the target range

While this objective has been a common consideration, researchers typically aim for RWL to be in close proximity to normal high water level (NHWL) or within a specified target range (Delgoda *et al.* 2013). Practical operators also seek to avoid RWL exceeding a certain water level to prepare for unexpected extreme events and maintain safety perceptions. Without this objective, optimal outflows often result in RWL approaching flood water level (FWL).

$$\tilde{J}5 := \min_{\mathbf{u}^k} \sum_{t=k}^{k+H-1} \delta S_U \times w_{S_U} + \delta S_L \times w_{S_L} + \delta S_H \times w_{S_H},$$

$$\delta S_U = \begin{cases} S_t - S_U & \text{if } S_t > S_U, \\ 0 & \text{otherwise}, \end{cases}$$

$$\delta S_L = \begin{cases} S_L - S_t & \text{if } S_L > S_t, \\ 0 & \text{otherwise}, \end{cases}$$

$$\delta S_H = \begin{cases} S_t - S_H & \text{if } S_t > S_H, \\ 0 & \text{otherwise}, \end{cases} \tag{5}$$

where $S_t$ represents the reservoir storage at time $t$, and $S_U$ and $S_L$ denote the storage levels for the upper and lower boundaries of the target range, respectively. The parameter $S_H$ corresponds to the highest water level that operators aim to avoid exceeding. We assign weights, denoted as $w_{S_U}$, $w_{S_L}$, and $w_{S_H}$, to these components. Generally, $S_U$ and $S_L$ have well-defined values that are less influenced by preference. $S_U$ can be NHWL or the restricted water level during the flood season, while $S_L$ may refer to the marginal storage reserved for water supply following a flood event. However, it is worth noting that $S_H$ is often subject to the operator's perceived risk tolerance with respect to exceeding FWL.

## 2.2.5. Continuity of spillway gates condition

One well-known limitation of MPC is that it can generate myopic control inputs when using a short prediction horizon (Morari & Lee 1999). In the context of reservoir flood control, this often leads to frequent opening and closing operations of spillway gates. This is not what operators usually want since it can increase the chance of actuators wearing out and human errors. Operators typically want to preserve the condition of spillway gates, i.e., when gates are already open, operators prefer not to close them until a flood event is certainly coming to an end. This objective can be formulated as follows:

$$\tilde{J}6 := \min_{\mathbf{u}^k} \sum_{t=k}^{k+H-1} \rho_t,$$

$$\rho_t = \begin{cases} 1 & \text{if } O_{spill,t-1}^k \times O_{spill,t}^k = 0 \text{ and } O_{spill,t-1}^k + O_{spill,t}^k > 0, \\ 0 & \text{otherwise}, \end{cases} \tag{6}$$

where $\rho$ is a variable to check if spillway gates are operated. When the gate condition changes, $\rho$ becomes 1. For example, when $O_{spill,t-1}$ is zero and $O_{spill,t}$ is not zero, then $O_{spill,t-1}^k + O_{spill,t}^k > 0$ and $O_{spill,t-1}^k \times O_{spill,t}^k = 0$. If the gate condition does not change, for example, both previous spillway outflow and current spillway outflow are zero, i.e., $O_{spill,t-1}^k + O_{spill,t}^k = 0$, or have positive values, i.e., $O_{spill,t-1}^k \times O_{spill,t}^k > 0$, and then $\rho$ becomes zero.

This equation shares a similar purpose to $\tilde{J}3$ and $\tilde{J}4$, in that it penalises changes in outflow schedules. However, it holds a greater significance because it also regulates the opening and closing of gates. The reason is that $\tilde{J}3$ and $\tilde{J}4$, which penalise total discharge, can have the minimum zero values while both turbine and spillway gate states change in a schedule. Especially, $\tilde{J}4$ has zero values when the change of spillway gate states is planned in a previous time step. Adding the objective shown in $\tilde{J}6$ will ensure that changes in spillway gates are specifically avoided unless necessary. However, this objective is a penalty form of what is called a complementarity constraint (Powell *et al.* 2016), so linearising this objective is difficult for MPC without adding binary variables representing $\rho$ resulting in a mixed-integer problem (Anitescu 2000). In Section 2.4, this objective is included only in the Evaluator to get around this issue of optimisation problem complexity.

### 2.2.6. Maintaining peak outflow under the peak inflow

Retaining inflow is one of the primary functions of reservoirs for flood control. Therefore, maintaining peak outflow under the peak inflow up to the current time step should be considered important. This can generally be achieved by $\bar{J}1$. However, it is also worth considering this by adding another objective as follows:

$$\bar{J}7 := \min_{\mathbf{u}^k} \gamma^k,$$
$$\gamma^k = \begin{cases} \text{large value} & \text{if } \max_{\forall\, t \in \{k,\dots,k+H-1\}} O_{total,t}^k > \max_{\forall\, t \in \{0,\dots,k-1\}} I_t, \\ 0 & \text{otherwise,} \end{cases} \tag{7}$$

where $I_t$ is the predicted inflow at time $t$.

Note that the peak outflow represents the maximum outflow within a prediction horizon, whereas the peak inflow denotes the maximum inflow up to the current time step, as shown in Equation (7). This objective can be easily linearised, but we incorporate it into the Evaluator in Section 2.4 rather than the MPC formulation because it is optimised indirectly in the middle of minimising the peak outflow in $\bar{J}1$ in most cases.

### 2.2.7. Soft constraint on utilising turbine capacity prior to opening spillway gate

To ensure efficient utilisation of turbine capacity before resorting to opening the spillway gate, a constraint is formulated as follows:

$$O_{spill,t}^k \times (O_{total,t}^k - MO_{turb}) = 0, \tag{8}$$

where $MO_{turb}$ is the turbine capacity. This constraint is also reformulated as a soft constraint, serving as an objective to circumvent the complexities of a nonlinear constraint:

$$\bar{J}8 := \min_{\mathbf{u}^k} \lambda^k,$$
$$\lambda^k = \begin{cases} \text{large value} & \text{if } O_{spill,t}^k > 0 \text{ and } O_{total,t}^k <= MO_{turb} \\ 0 & \text{otherwise.} \end{cases} \tag{9}$$

This is also a complementarity constraint, which is hard to linearise. Therefore, in this research, we include this constraint as an objective in the evaluator using a penalty approach for the same reason as $\bar{J}6$.

### 2.3. The dynamical characteristic of the operator's preferences

As mentioned in Section 1, numerous studies aimed to find the best set of objective weights while assuming that the preference remains constant. This is because the relative importance of each objective typically does not change significantly when hydrological conditions remain stable. However, in the context of reservoir flood control, operating preference can often shift with changes in hydrological states. This variability is evident from the objectives outlined in Section 2.2.

Some objectives may conflict with one another, while others may complement each other, and this dynamic depends on the current state. For instance, when RWL approaches FWL or $S_H$, the importance of $\bar{J}5$ increases, necessitating a substantial increase in outflows with less importance for other objectives. Conversely, if RWL remains below $S_U$ and spillway outflows are stable, $\bar{J}3$ and $\bar{J}4$ should be prioritised. When a significant increase in outflows becomes unavoidable, the highest weight should be assigned to $\bar{J}1$, followed by $\bar{J}5$. Consequently, we can conclude that the relative importance of each objective should vary depending on hydrological conditions.

Some parameters, like the target water level, have been considered static as well. However, the parameters such as $S_U$, $S_L$, and $S_H$ in Equation (5) directly impact the objective value of $\bar{J}5$ and the optimal control inputs. Hence, we can not assume that these parameters remain static when the preference is not static.

## 2.4. Parameterised dynamic MPC framework

MPC is widely used in reservoir flood control, leveraging system models to predict future states and determine optimal control inputs (Breckpot *et al.* 2013; Delgoda *et al.* 2013; Castelletti *et al.* 2023). When we assume dynamically changing objectives' weights and parameters, both control inputs and objectives' weights/parameters need to be optimised simultaneously. The problem becomes complicated due to their interdependency. Without additional criteria, this co-optimisation problem would be intractable.

If we have the means to evaluate the derived optimal control inputs, which we call an evaluator here, we can simplify this problem. While traditional approaches typically generate a Pareto set, a set of alternatives, and select the optimal one using a single criterion, such as the possibility for system failure (Chen *et al.* 2020), minimising the worst-case impact (Yu *et al.* 2023), or integrate simple but diverse evaluation criteria (Zhu *et al.* 2018; Myo Lin *et al.* 2020), computing the complete Pareto set becomes computationally intractable with multiple objectives (eight in our case), particularly for the online optimisation process of MPC.

To circumvent this complexity, we suggest a framework for searching for the best weights/parameters that optimise many objectives and satisfy constraints, accounting for the dynamic hydrological state. Instead of finding a Pareto set, this framework simultaneously searches the weight and parameter spaces on the basis of the evaluator in order to produce the optimal control inputs. We refer to this framework as parameterised dynamic model predictive control (PD-MPC) as illustrated in Figure 1. Here, the disturbance, denoted as $\boldsymbol{d}_p^k$, represents an uncertain inflow vector at time step $k$, i.e., for time $k$ to $k + H - 1$, where $H$ is the length of the prediction horizon. The state vector at time step $k$ is denoted as $\mathbf{x}^k$. $\mathbf{u}$ refers to the control inputs for a given weight and parameter vector, $\mathbf{z}$. $E$ denotes the absolute performance evaluator.

The PD-MPC approach solves the optimal control problem at the $k$th step of a receding horizon implementation:

$$\{\mathbf{z}^{k*}, \mathbf{u}^{k*}\} = \arg \min_{\mathbf{z}^k, \mathbf{u}^k, \mathbf{x}^k} E(\mathbf{z}^k, \mathbf{u}^k, \mathbf{x}^k | \mathbf{x}_0^k), \tag{10}$$

by alternatingly solving the two sets of problems:

$$\{\mathbf{z}^{k*}\} = \arg \min_{\mathbf{z}} \mathbf{E}(\mathbf{z} | \mathbf{x}^k, \mathbf{u}^k), \text{ and} \tag{11a}$$

$$\{\mathbf{u}^k | \mathbf{z}^k\} = \arg \min_{\mathbf{x}^k, \mathbf{u}^k} \mathbf{J}(\mathbf{x}^k, \mathbf{u}^k | \mathbf{x}_0^k, \mathbf{z}^k), \tag{11b}$$

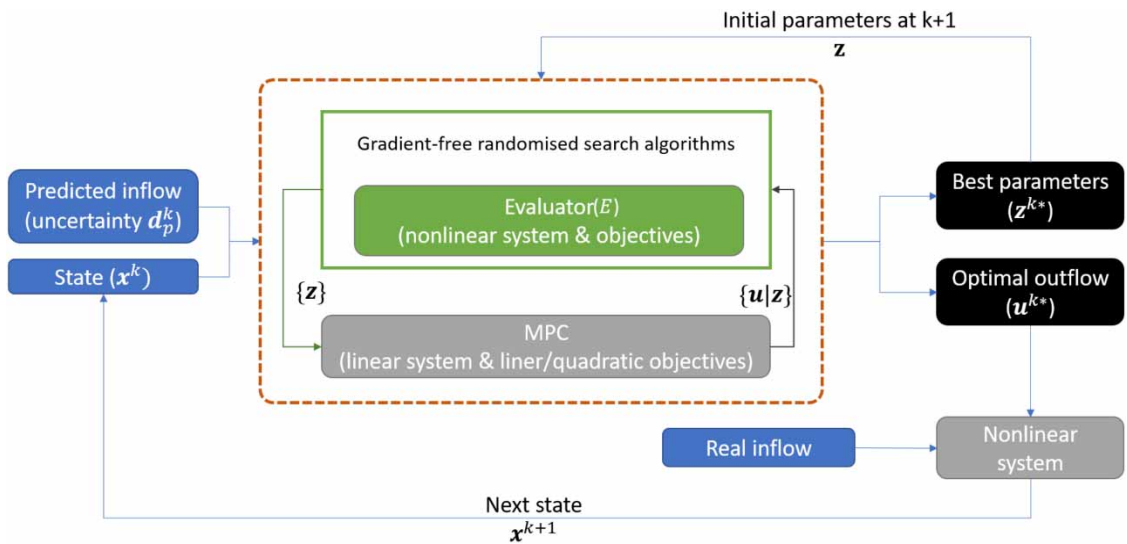

**Figure 1** | The schematic diagram of PD-MPC approach, with $\mathbf{z}^k$ warm started with optimal values from the previous step.

respectively, where $\mathbf{x}_0^k$ represents the initial state at time $k$, $\mathbf{J}(\cdot)$ is the objectives of MPC. In this equation, $E(\cdot)$ is not directly optimised; instead, it just evaluates $\mathbf{z}|\mathbf{x}^k$, $\mathbf{u}^k$, thus, $E(\cdot)$ can be formulated using highly nonlinear equations, e.g., conditional equations and exponential formulas, without limitation.

At each time step, we first solve the Equation (11b) to obtain control actions for a specific weight set, then solve the Equation (11a) to optimise both the weights/parameters and their corresponding control inputs through the evaluator, $E(\cdot)$. To solve this evaluation problem, we can employ heuristic optimisation techniques, such as GA. Heuristic algorithms explore the weight and parameter space to find weights and parameters with the optimal value from the evaluator at each time step of the receding horizon recalculation. Therefore, this methodology enables us to decouple the optimisation process into two separate tasks: finding optimal weights and determining control actions, which simplifies the co-design problem, as illustrated in Figure 1.

In this research, we linearise all proposed practical objectives for a linear MPC and integrate objectives presented in the previous section to formulate the evaluator, without limiting the evaluator to linear and nonlinear objectives that are straightforward to implement in mathematical optimisation. Moreover, to reflect real hydrological processes, the evaluator uses a nonlinear system simulation (i.e., nonlinear height-volume curve) to calculate water levels. This allows the PD-MPC framework, primarily optimising a linear MPC problem to find a release schedule, to select an outflow schedule that actively aligns with the operator's preferences. The reason why all objectives are linearised for $\mathbf{J}(\cdot)$, instead of applying quadratic equations as presented in Section 2.2, is to penalise even minor exceedances and linearising an absolute form is straightforward. For detailed explanations, including formulas, please refer to Supplementary material, Appendix A for the evaluator and Section 3.3 for the objectives, weights and parameters.

In addition, we utilise GA as our heuristic optimisation algorithm. The genetic algorithm (GA) is one of the most prominent and easy-to-implement gradient-free algorithms which imitate the natural selection process (see, e.g., Katoch *et al.* 2021). It continuously develops the population (a set of vectors seen as potential solutions) over each iteration, employing reproduction, crossover, and mutation operators, aiming at preserving the vectors with lower objective values and iteratively recombining them.

## 3. NUMERICAL EXPERIMENT

### 3.1. Description of the case study area

The Daecheong reservoir is located in the middle of South Korea, within the upper basin of the Geum river, as illustrated in Figure 2. There is another multipurpose reservoir, the Yongdam reservoir, located in the upper basin of the Daecheong reservoir. The outflows from the Yongdam reservoir have a direct influence on the flood control strategies of the Daecheong reservoir since they enter the Daecheong reservoir basin directly. Usually and understandably, operators decide on the outflows of the Yongdam reservoir first, followed by decisions on the outflows of the Daecheong reservoir. Morever, the outflow from Yongdom does not significantly affect the Daecheong reservoir because it has a great amount of storage capacity and discharges only a small amount during a flood event in general. Therefore, we assume that we have prior knowledge of the future outflows from the Yongdam reservoir for simplicity. The specifications of the Daecheong reservoir are presented in Table 1.

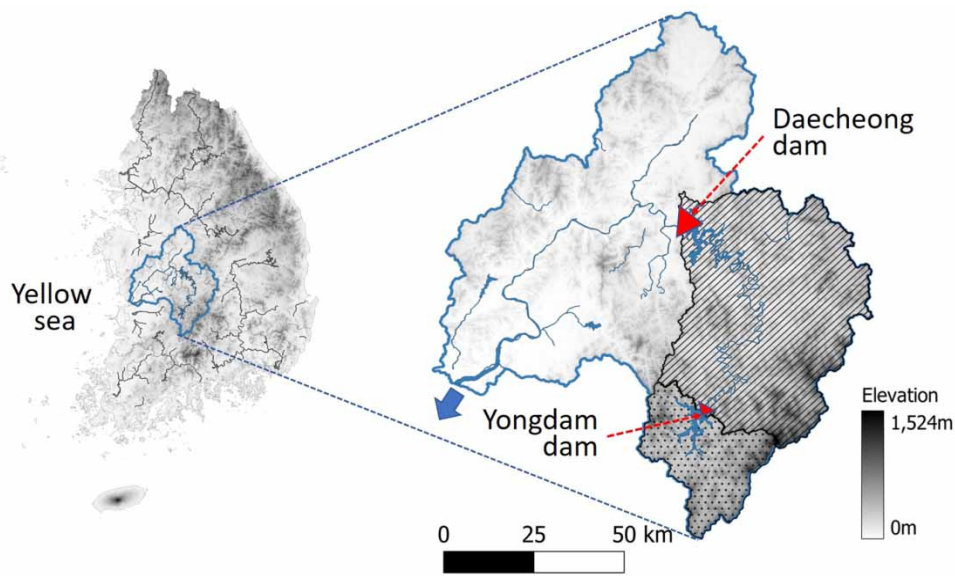

**Figure 2** | The Daecheong Reservoir, illustrated in the right panel, is situated upstream in the Geum River basin, located in central South Korea, as depicted in the left panel. In the right figure, red triangles indicate dam structures, and the Geum River flows from the bottom right to the middle left, ultimately discharging into the Yellow Sea.

The Daecheong reservoir has enough spillway capacity with the spillway crest level significantly lower than NHWL. The outflow capacity at NHWL is greater than 6,000 m³/s, so we do not need to consider the spillway outflow capacity according to RWL. However, the flood control storage is relatively small. For example, the 200-year frequency inflow is 10,700 m³/s, and it takes only 6.5 h to completely fill the flood control storage. In addition, there is no restricted water level for the flood season in this reservoir. In essence, the Daecheong reservoir is susceptible to flooding, necessitating rapid and well-informed flood control decisions by operators to avert extreme conditions.

Three flood events with a large amount of peak inflow and one to three peaks are selected from historical data as presented in Table 2. To the best of our knowledge, most studies have primarily concentrated on floods lasting 1–2 days (Uysal et al. 2018a; Breckpot et al. 2013; Delgoda et al. 2013). In the case of short flood events, generating reliable outflows is relatively straightforward because the trade-offs among objectives are evident. For instance, it is clear that minimising outflows leads to an increase in reservoir water level, and vice versa. However, the situation becomes more complex over longer time frames. Minimising outflows can ironically result in increased outflows if peak inflow occurs after the reservoir water level has already risen significantly. Therefore, our research examines floods lasting more than eight days, featuring one to three peaks, to provide a more practical case study.

Although these events can present significant operational challenges demonstrating our method's practical utility, more extreme floods, such as a 200-year return period flood and a probable maximum flood (PMF), would provide more rigorous validation cases. However, the Daecheong reservoir, our case study site, was constructed over 40 years ago, and to the best of our knowledge, the hydrographs for the 200-year flood and PMF are unavailable. Moreover, the 200-year flood and PMF typically span only 1–3 days with a single peak, representing less uncertainty and total inflow volume.

To simulate the uncertainty associated with inflow predictions, we apply a conventional method in which uncertainty follows a Gaussian distribution with zero means (Uysal et al. 2018a; Li et al. 2010), instead of an ensemble or other scenario generation methods, for simplicity. Thus, 'predicted' inflow data is generated by multiplying random numbers by real inflow, as Equation (12). A standard deviation of random numbers is $a$. Predicted inflow should be positive; therefore, the lower boundary $c$ is introduced. Forecasting uncertainty tends to increase as the forecast period increases. To reflect this, we adjust the standard deviation by increasing it by $b$ per each time step. Moreover, a moving average filter with a 3-h window is applied every time step. This filter helps prevent unreasonable high fluctuations in predicted inflow within the prediction horizon.

**Table 1** | Specifications of the Daecheong reservoir

| | | | |
|---|---|---|---|
| Flood water level (FWL) | EL. 80.0 m | Total storage | 1,490 10⁶ m³ |
| Normal high water level (NHWL) | EL. 76.5 m | Spillway capacity | 11,680 m³/s |
| Low water level (LWL) | EL. 60.0 m | Turbine capacity | 264 m³/s |
| Spillway crest level | EL. 64.5 m | 200-year frequency flood | 10,700 m³/s |

**Table 2** | Study flood events

| Periods | Duration | Peak inflow | Feature |
|---|---|---|---|
| From 2 July 2016 to 11 July 2016 | 197 h | 3,655 m³/s | Double peak |
| From 25 August 2018 to 7 September 2018 | 311 h | 2,590 m³/s | Triple peak |
| From 1 September 2020 to 12 September 2020 | 269 h | 4,468 m³/s | One peak |

$$PI_t = \frac{1}{3} \sum_{j=t-2}^{t} RI_j \times \omega I_t,$$

$$\omega I_t = \begin{cases} 1 + x, & \text{where } x \sim \mathcal{N}(0, (a + T \times b)^2) & \text{if } 1 + x \geq c, \\ c & \text{otherwise,} \end{cases} \tag{12}$$

where $PI_t$ and $RI_t$ are the predicted and real inflow at time $t$, while $\omega I_t$ denotes the uncertainty of inflow at time $t$. The variable $T$ is the index of forecasting time $t$ within the current prediction horizon (i.e., $T = 0 : N - 1$), so the prediction uncertainty increases for a time farther into the future. The constants $a$, $b$, $c$ are set to 0.05, 0.03, and 0.1—somehow arbitrarily here—to generate something that has roughly the same order of errors as in reality; the sample hydrographs of real and predicted inflow provided in Supplementary material, Appendix B demonstrate visually this method is reliable.

## 3.2. A reservoir flood control system

### 3.2.1. System model

A simple linear reservoir model can be expressed as

$$S_{t+1} = S_t + I_t - O_{total,t}, \tag{13}$$

where $S_t$ is the current storage of a reservoir at time $t$, and $I_t$ and $O_t$ are the total amount of inflow and outflow for time $t$, respectively. Errors in measuring water level, calculating storage amount and outflow are not considered for simplicity. In addition, evaporation and seepages are not explicitly considered in Equation (13), but these factors are already accounted for in the inflow data. This is because inflow is not directly measured but calculated using Equation (13) from measured outflow and storage timeseries for the Daecheong reservoir.

### 3.2.2. Constraints

The operational constraints can be formulated as follows:

1. The total outflow is more than the sum of the agreed amounts with water users and the in-stream flow, i.e., the downstream demand:

$$D_t \leq O_{total,t}, \tag{14}$$

where $O_{total,t}$ and $D_t$ are the total outflows and the downstream demand at time $t$, respectively.
2. RWL should be between FWL and LWL:

$$LWS \leq S_t \leq FWS, \tag{15}$$

where $S_t$ is the reservoir storage at time $t$, while LWS and FWS are the reservoir storage at LWL and FWL, respectively.
3. Outflows via the turbine and spillway gates are not able to exceed their capacity:

$$O_{turb,t} \leq MO_{turb}, \\ O_{spill,t} \leq MO_{spill}, \tag{16}$$

where $O_{turb,t}$ and $O_{spill,t}$ represent the turbine and spillway outflow at time $t$, respectively. $MO_{turb}$ and $MO_{spill}$ are the outflow capacity of turbines and spillway gates. We take into account the spillway crest level in our analysis, but spillway outflow capacity depending on RWL is disregarded as explained in Section 3.1.

### 3.2.3. Objectives

By linearising the practical objectives described in Section 2.2 to penalise even minor exceedances, the objective function of MPC is set as below:

$$\min J1^k + J2^k + J3^k + J4^k + J5^k, \tag{17}$$

where,

$$J1^k := \max_{t \in k,...,k+H-1} O^k_{spill,t} \times w_1, \tag{18a}$$

$$J2^k := \sum_{t=k}^{k+H-1} O^k_{spill,t} \times w_2, \tag{18b}$$

$$J3^k := \sum_{t=k+1}^{k+H-1} \delta O^{in,I}_t \times w_{3,I} + \delta O^{in,D}_t \times w_{3,D}, \tag{18c}$$

$$J4^k := \sum_{t=k}^{k+H-1} \delta O^{between,I}_t \times w_{4,I} + \delta Q^{between,D}_t \times w_{4,D}, \tag{18d}$$

$$J5^k := \sum_{t=k}^{k+H-1} \mu S^1_t \times w_{5,1} + \mu S^2_t \times w_{5,2} + \mu S^3_t \times w_{5,3}, \tag{18e}$$

subject to,

$$O^k_{spill,t} + O^k_{turb,t} - O^k_{total,t} = 0, \tag{19a}$$

$$O^k_{total,k} - O^{k-1}_{total,k} = 0, \tag{19b}$$

$$\delta O^{in,I}_t - \delta O^{in,D}_t + (O^k_{total,t} - O^k_{total,t-1}) \times w_{\delta O_{in,t}} = 0, \tag{19c}$$

$$\delta O^{between,I}_t - \delta O^{between,D}_t + (O^k_{total,t} - O^{k-1}_{total,t}) \times w_{\delta O_{between,t}} = 0, \tag{19d}$$

$$\mu S^1_t + S_U - S_t \geq 0, \tag{19e}$$

$$\mu S^2_t + S_t - S_L \geq 0, \tag{19f}$$

$$\mu S^3_t + S_H - S_t \geq 0, \tag{19g}$$

$$\delta O^{in,I}_t, \delta O^{in,D}_t, \delta O^{between,I}_t, \delta O^{between,D}_t, \mu S^1_t, \mu S^2_t, \mu S^3_t \geq 0, \tag{19h}$$

where $J^k_1$, for example, is an objective for the MPC formulation at time step $k$. All objectives are multiplied by their respective weights and then summed together to form the objective function, as shown in Equations (17) and (18). Weights, i.e., $w_1$, $w_2$, $w_{5,I}$, $w_{5,D}$, $w_{4,I}$, $w_{4,D}$, $w_{5,1}$, $w_{5,2}$, and $w_{5,3}$, are multiplied by each objective.

Several positive slack variables are introduced to linearise objectives instead of applying absolute forms. These variables include $\delta O^{in,I}_t$ and $\delta O^{in,D}_t$, representing the step-wise increase and decrease in outflows in a prediction horizon; $\delta O^{between,I}_t$ and $\delta O^{between,D}_t$, representing the increase and decrease in outflow schedules at time step $k$ and $k-1$; $\mu S^1_t$, $\mu S^2_t$, and $\mu S^3_t$, representing the storage violation above $S_U$, below $S_L$ and above $S_H$, respectively. To account for varying penalties over time, we assign weights $w_{\delta O_{in,t}}$ and $w_{\delta O_{between,t}}$ to each change at time $t$. The idea is to penalise changes closer to the start of the horizon more than changes farther ahead in the horizon. In this work, we set $w_{\delta O_{in,t}}$ to $\frac{1}{(t-k+1)\times 3}$, and $w_{\delta O_{between,t}}$ is set as follows:

$$w_{\delta O_{between,t}} = \begin{cases} \dfrac{1}{t-k+1} & \text{if } t \leq k+3, \\ \dfrac{1}{(t-k+1)\times 2} & \text{otherwise.} \end{cases} \tag{20}$$

The target water levels, such as $S_U$ and $S_L$, are defined as mentioned in Section 2.2. The weights of objectives in Equation (18), as well as the allowed highest water level, $S_H$ in Equation (19g), will be found during the optimisation process.

The important difference in the objective formulas presented in Section 2.2 is about $O^k_{total,k}$ in Equation (19b) and $\tilde{J}4$ in Equation (4). Note that $O^k_{total,k}$ is the first outflow in the optimal control inputs at time step $k$. First, changing the outflow for time $k$ at time step $k$ is practically impossible because it involves implementing the current outflow while calculating it. To address this, we separate this to ensure it, rather than assigning significant weight to the change in the first outflow in Equation (19d). Hereby, the outflow decision is delayed for a time step.

### 3.3. PD-MPC design

Since the Korea Meteorological Agency (KMA) publishes 6-h quantitative rainfall forecasts every hour, in the experiment, we employed four different prediction horizons: 6, 12, 18 and 24 h. Starting from the 6 h, the longer horizons allow us to explore the effect of horizon length on performance. The control horizon is the same as the prediction horizon. For the initial run at $t = 0$, we set the initial storage to the corresponding level of NHWL (EL. 76.5 m), while the initial outflows via the turbines and spillway gates are set to $150 \, \mathrm{m^3/s}$, which is the average hourly outflow during the flood season from 2015 to 2020, and $0 \, \mathrm{m^3/s}$, respectively.

In this experiment, the GA optimises 10 parameters, comprising of 9 weights for the objectives and 1 parameter for the highest RWL, denoted as $S_H$. To reduce the running time of GA, we impose search range limits for each weight and parameter, as outlined in Table 3. In $J5$ in Equation (18), $w_{5,2}$ and $w_{5,3}$ are fixed at twice and twenty times $w_{5,1}$, respectively, because maintaining RWL under $S_H$ for dam safety is a higher priority than maintaining it between target levels, additionally, in order to reduce the computational complexity by decreasing the number of weights which need to be optimised. Again, the reason why nominators for $w_{5,2}$ and $w_{5,3}$ are larger than others is to reduce the number of dynamic weights and to reflect the fact that the objectives corresponding to $w_{5,2}$ and $w_{5,3}$ are more important than the objective for $w_{5,1}$. Here, 40 and 400 are selected somewhat arbitrarily because precise values are not critical due to the PD-MPC, which finds optimal weights dynamically. Similar to the value of 20 for other multipliers, the values 40 and 400 are selected to prevent the solver from neglecting objectives when objective values are divided by a substantial value, e.g., FWS for $w_{5,1}$, $w_{5,2}$, and $w_{5,3}$. Instead of fixing these nominators and weights, we could extend the search range from 1–20 to 1–400 or 1–500, and so on. However, the results are the same, although it takes considerably more time. This is why we fixed the nominators instead of extending the search range while setting the same value for all nominators. Therefore, the number of weights and parameters explored becomes 8 from 10. All weights share the same length of the search range, except for $J2$. This is because the weight of $J2$, i.e., minimising the total outflows via spillway gates, can be relatively small due to its similarity to $J1$, i.e., minimising the peak outflows via spillway gates, and we want to emphasise $J1$. To ensure $J5$ is not neglected, the search ranges for $w_{5,1}$, $w_{5,2}$, and $w_{5,3}$ start from 1, not 0.

Moreover, to simplify the optimisation process and work with integer values in GA, we introduce multipliers based on these MAVEs. Because MAVE represents the largest anticipated magnitude of each objective, it is often adopted to normalise objectives to maintain numerical balance across different objectives. Each weight is then calculated by multiplying a selected integer value by its corresponding multiplier. This approach ensures that all objectives contribute as a proportion of the selected values in the search range, regardless of their natural scales, while allowing GA to work with simpler integer values during the search process.

We prepare MPC baselines with fixed weights and a fixed parameter set, denoted as 'Fixed', which would be used in standard operation, as presented in Table 3. Our comparative analysis aims to demonstrate that operators' decisions can be improved by dynamically optimising weights in real-time. This approach allows for adapting risk preferences depending on the system state. To ensure the feasibility of the problem, two weight/parameter sets, which emphasise minimising the

**Table 3** | The possible range for PD-MPC and fixed weights/parameters for the baseline MPC

| P | $w_1$ | $w_2$ | $w_{3,J}$ | $w_{3,D}$ | $w_{4,J}$ | $w_{4,D}$ | $w_{5,1}$ | $w_{5,2}$ | $w_{5,3}$ | $S_H$ |
|---|---|---|---|---|---|---|---|---|---|---|
| Multiplier | $\frac{20}{MO_{spill}}$ | $\frac{2}{MO_{spill}}$ | $\frac{20}{MO_{spill}}$ | $\frac{20}{MO_{spill}}$ | $\frac{20}{MO_{spill}}$ | $\frac{20}{MO_{spill}}$ | $\frac{20}{FWS \times F}$ | $\frac{40}{FWS \times F}$ | $\frac{400}{FWS \times F}$ | – |
| Search range | 0–19 | 0–2 | 0–19 | 0–19 | 0–19 | 0–19 | 1–20 | ← | ← | $RWS$ |
| Fixed-1 | 3 | 1 | 3 | 3 | 20 | 20 | 15 | ← | ← | $RWS_F$ |
| Fixed-2 | 20 | 5 | 3 | 3 | 3 | 3 | 15 | ← | ← | $RWS_F$ |

*Note* [1] Multipliers are introduced to normalise objective values using MAVE as the denominator of each objective, and nominators are set to prevent objective values from approaching zero in any case as well as to reduce the range of search space.

[2] ← indicates the same value as in the previous column.

[3] Fixed-1 and Fixed-2 represent MPC baselines with fixed weights and a fixed parameter set.

changes in outflow schedules and minimising peak outflow, are selected from the best weight/parameter sets generated from various PD-MPC tests and verified through trial and error (Uysal *et al.* 2018a). Throughout this paper, we refer to the value before being multiplied by the multiplier as the weights unless this would cause confusion otherwise.

In Table 3, a searching range of $S_H$ is the storage where RWL is in {EL. 78.5 m, 79.0, 79.5 m} and $S_H$ of the baselines, i.e., $RWS_F$ (Reservoir Water Storage for Fixed weights/parameter set), which is the stored amount of water at EL.79.0 m. A computer code is developed using Python. In detail, pyomo (Hart *et al.* 2011), GLPK solver (Makhorin n.d) and pyGAD (Gad 2021) packages were applied to implement the numerical experiment of PD-MPC.

## 4. RESULTS AND DISCUSSION

### 4.1. PD-MPC results

Parameterised dynamic model predictive control (PD-MPC) delivers reliable results across all events and prediction horizons. The results show only a few changes in outflow schedules, and all peak outflows remain below the peak inflow.

Figure 3 illustrates the optimal outflows with uncertain inflow and RWL for Event 1 across different prediction horizons. The generated optimal outflows, denoted as $\mathbf{u}^k$s, are represented by dashed lines of various colors, overlaid with the red line indicating the implemented total outflows in the figures. For instance, in Figure 3(d), where outflows are decreasing, a 24-h prediction horizon leads to numerous changes in outflows (see also Table 4). The figures show only a few changes in outflow schedules, and all peak outflows remain below the peak inflow.

The detailed result of the numerical experiment is presented in Table 4 for uncertain inflow and Table 5 for certain inflow. As expected, the performance of MPC under certainty generally surpasses the results obtained under uncertainty. Peak outflows and RWLs from uncertain inflow exceed those from certain inflow, and the reservoir should change the outflow schedule more often. Additionally, under uncertain inflow conditions for Event 1, a comparison of the results between PD-MPC and the Fixed cases is presented in Table 6. PD-MPC outperforms the Fixed cases for all items in the table.

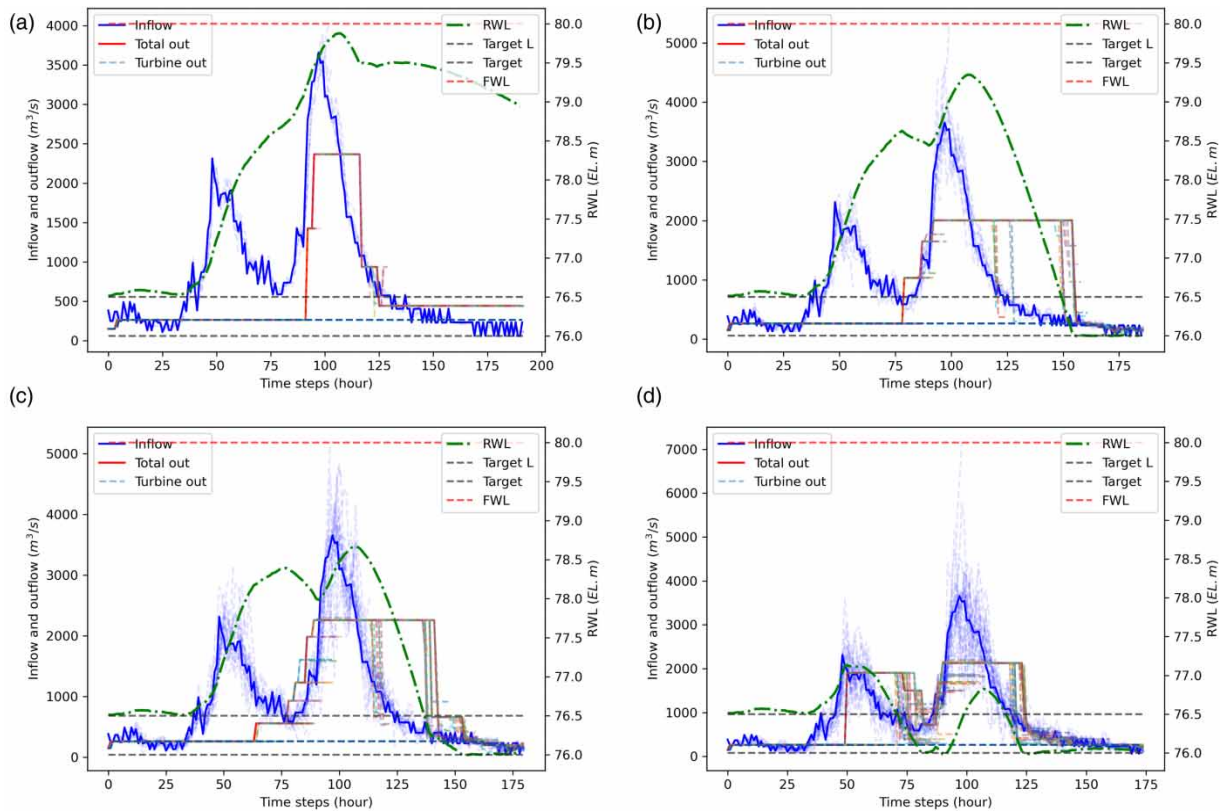

**Figure 3** | Hydrographs under uncertain inflow and different prediction horizons for Event 1. (a) 6 h; (b) 12 h; (c) 18 h; and (d) 24 h.

**Table 4** | The detailed result of PD-MPC under uncertain inflow

| Prediction horizon | Event | Peak outflow (m³/s) | Peak RWL (EL. m) | Lowest RWL (EL. m) | Changes between time steps |
|---|---|---|---|---|---|
| 6 | 1 | 2,367 | 79.88 | 76.50 | 9 |
| 12 | 1 | 2,005 | 79.35 | 76.00 | 9 |
| 18 | 1 | 2,296 | 78.67 | 76.00 | 11 |
| 24 | 1 | 2,128 | 77.13 | 75.98 | 9 |
| 6 | 2 | 1,919 | 79.82 | 76.13 | 20 |
| 12 | 2 | 1,591 | 79.04 | 76.00 | 8 |
| 18 | 2 | 1,273 | 77.56 | 76.00 | 21 |
| 24 | 2 | 1,915 | 76.81 | 76.01 | 30 |
| 6 | 3 | 2,105 | 79.81 | 76.34 | 13 |
| 12 | 3 | 1,323 | 78.37 | 75.99 | 10 |
| 18 | 3 | 1,245 | 78.19 | 76.00 | 18 |
| 24 | 3 | 1,273 | 78.07 | 76.01 | 38 |

**Table 5** | The detailed result of PD-MPC under certain inflow

| Prediction horizon | Event | Peak outflow (m³/s) | Peak RWL (EL. m) | Lowest RWL (EL. m) | Changes between time steps |
|---|---|---|---|---|---|
| 6 | 1 | 2,332 | 79.89 | 76.50 | 9 |
| 12 | 1 | 2,153 | 79.27 | 76.00 | 4 |
| 18 | 1 | 2,290 | 78.60 | 76.00 | 11 |
| 24 | 1 | 1,873 | 77.13 | 75.93 | 4 |
| 6 | 2 | 2,051 | 79.84 | 76.19 | 9 |
| 12 | 2 | 1,499 | 79.09 | 75.99 | 8 |
| 18 | 2 | 990 | 77.38 | 76.00 | 11 |
| 24 | 2 | 1,870 | 76.92 | 76.00 | 17 |
| 6 | 3 | 2,078 | 79.82 | 76.32 | 8 |
| 12 | 3 | 1,051 | 78.59 | 76.00 | 11 |
| 18 | 3 | 1,104 | 78.33 | 75.95 | 9 |
| 24 | 3 | 1,374 | 77.91 | 76.00 | 18 |

The performance of this framework can be presented clearly when compared to the Fixed cases, as shown in Figure 4 under uncertain inflow. The Fixed-1 case assigns a relatively high weight to minimise the changes in outflow schedules, while the Fixed-2 case prioritises minimising peak outflow, as illustrated in Table 3. The results challenge our expectations. The Fixed cases do not consistently outperform PD-MPC, even in their main targets. It is impossible to assert that the Fixed-1 case having a high weight for minimising the changes in outflow schedules is always superior to the other two cases, PD-MPC and Fixed-2, even concerning the changes in outflow schedules, as shown in Figure 4(a) and 4(d). Similarly, peak outflows of the Fixed-2 case are not certainly less than the other two cases, as demonstrated in Figure 4(b) and 4(e), even though it has a high weight for minimising peak outflow compared to the other cases. This suggests that fixed weights fail to consistently represent the operators' preferences. This can be attributed to the complexity of trade-offs for each time step and the relatively shorter prediction horizon length compared to the length of a flood event. For example, we often assume that a lower peak outflow leads to a higher peak RWL. However, the results show that this assumption does not always hold true, as demonstrated in Figure 4(c) and 4(f). This is because the *myopic* outlook on optimal outflows at each time step could lead to suboptimal outcomes when viewed in the context of the entire flood event (Morari & Lee 1999; Bøhn *et al.* 2021). In our case, the evaluator effectively makes MPC have a long-term perspective compared to the Fixed weight cases.

**Table 6** | The comparison between PD-MPC and the Fixed cases for Event 1 under uncertain inflow

|  | Prediction horizon | Peak outflow (m³/s) | Peak RWL (EL. m) | Lowest RWL (EL. m) | Changes between time steps |
|---|---|---|---|---|---|
| PD-MPC | 6 | 2,367 | 79.88 | 76.50 | 9 |
| Fixed 1 | 6 | 3,710 | 79.10 | 75.97 | 13 |
| Fixed 2 | 6 | 3,617 | 79.31 | 76.31 | 28 |
| PD-MPC | 12 | 2,005 | 79.35 | 76.00 | 9 |
| Fixed 1 | 12 | 3,267 | 79.19 | 76.37 | 10 |
| Fixed 2 | 12 | 2,759 | 79.35 | 76.34 | 37 |
| PD-MPC | 18 | 2,259 | 78.67 | 76.00 | 11 |
| Fixed 1 | 18 | 2,578 | 79.35 | 76.34 | 12 |
| Fixed 2 | 18 | 2,551 | 79.60 | 76.31 | 48 |
| PD-MPC | 24 | 2,128 | 77.13 | 75.98 | 9 |
| Fixed 1 | 24 | 2,259 | 78.03 | 75.95 | 16 |
| Fixed 2 | 24 | 2,223 | 79.32 | 76.28 | 33 |

In addition, PD-MPC shows smaller peak outflow as well as lower peak RWL, especially in Figure 4(b) and 4(c). This implies that PD-MPC utilised the storage capacity more effectively than the Fixed cases for this flood event.

In Figure 5(a), the maximum penalty value of PD-MPC is less than 40 in any case. Additionally, the sums of penalty values for the entire flood event using PD-MPC across all prediction horizons show the lowest values. In Section 2.2, we described undesirable conditions, such as the peak outflow exceeds the maximum inflow up to the current time steps because retaining inflow is one of the reservoir's fundamental roles for flood control, RWL reaches FWL. We can see this also in Figure 5(a). Since we assigned the 'large value' as 1,000 in Equation (7) in Section 2.2, under normal operating conditions (i.e., without these undesirable conditions), the maximum penalty values are constrained below 1,000. Notably, both fixed cases exhibit these undesirable conditions many times, contrary to the fact that it has never occurred in PD-MPC.

To see the impact of the evaluator, we systematically adjusted the importance of the changes in outflows calculated at consecutive time steps and compared it with the previous one. The result is straightforward. Figures 6–8 show the effect of the evaluator by changing the weight of the objective, which minimises the changes in outflow schedules, i.e., $\tilde{J}4$ in Section 2.2, in the evaluator. When the evaluator assigns a higher weight to $\tilde{J}4$ marked as 'Higher' in figures, it leads to fewer changes in outflows but comes at the expense of peak outflow and RWL. PD-MPC with the low $\tilde{J}4$ case marked as 'Lower' in figures shows a greater number of changes. Given the unambiguous response from PD-MPC, it can be inferred that the evaluator effectively affects the selection of the optimal weight/parameter set.

## 4.2. Parameters as elements of operators' preference

In Section 1, we discussed that some researchers had focused on finding appropriate weights of objectives when defining preferences in a multi-objective setting. To demonstrate the parameters should also be regarded as important components of the preference, we conducted an experiment for Event 1, where we fixed $S_H$ to the storage level at EL. 79.0 m and compare it with the results of PD-MPC where $S_H$ is a dynamic parameter to be optimised.

PD-MPC with varying $S_H$ outperforms PD-MPC with fixed $S_H$, which generally considers only the combination of weights as the representative of the preference, as illustrated in Figure 9. PD-MPC shows low peak outflows and RWL with the lowest number of changes simultaneously. It seems reasonable to assume that operators would prefer the evaluator to have a lower $S_H$ because it results in a lower peak RWL. However, this is not the case because a lower $S_H$ can lead to a higher peak outflow. PD-MPC, by varying $S_H$, effectively achieves lower peak outflows and RWLs simultaneously. This means an adaptively changing $S_H$ helps operators utilise the reservoir storage more efficiently. For instance, in PD-MPC with the parameter $S_H$, spillway outflow tends to commence earlier compared to PD-MPC without $S_H$, achieved by maintaining a low $S_H$ before the water level rises significantly. Consequently, there is an impact on reducing both the peak outflow and the peak RWL. In addition, when RWL is close to $S_H$, and there is a sudden increase in inflow, fixing $S_H$ would result in a substantial and abrupt increase in outflow, even if there is some storage available between $S_H$ and FWL. In contrast, with adaptively changing $S_H$, PD-MPC chooses to increase $S_H$ to utilise the remaining storage, instead of resorting to a drastic and sudden increase in outflow.

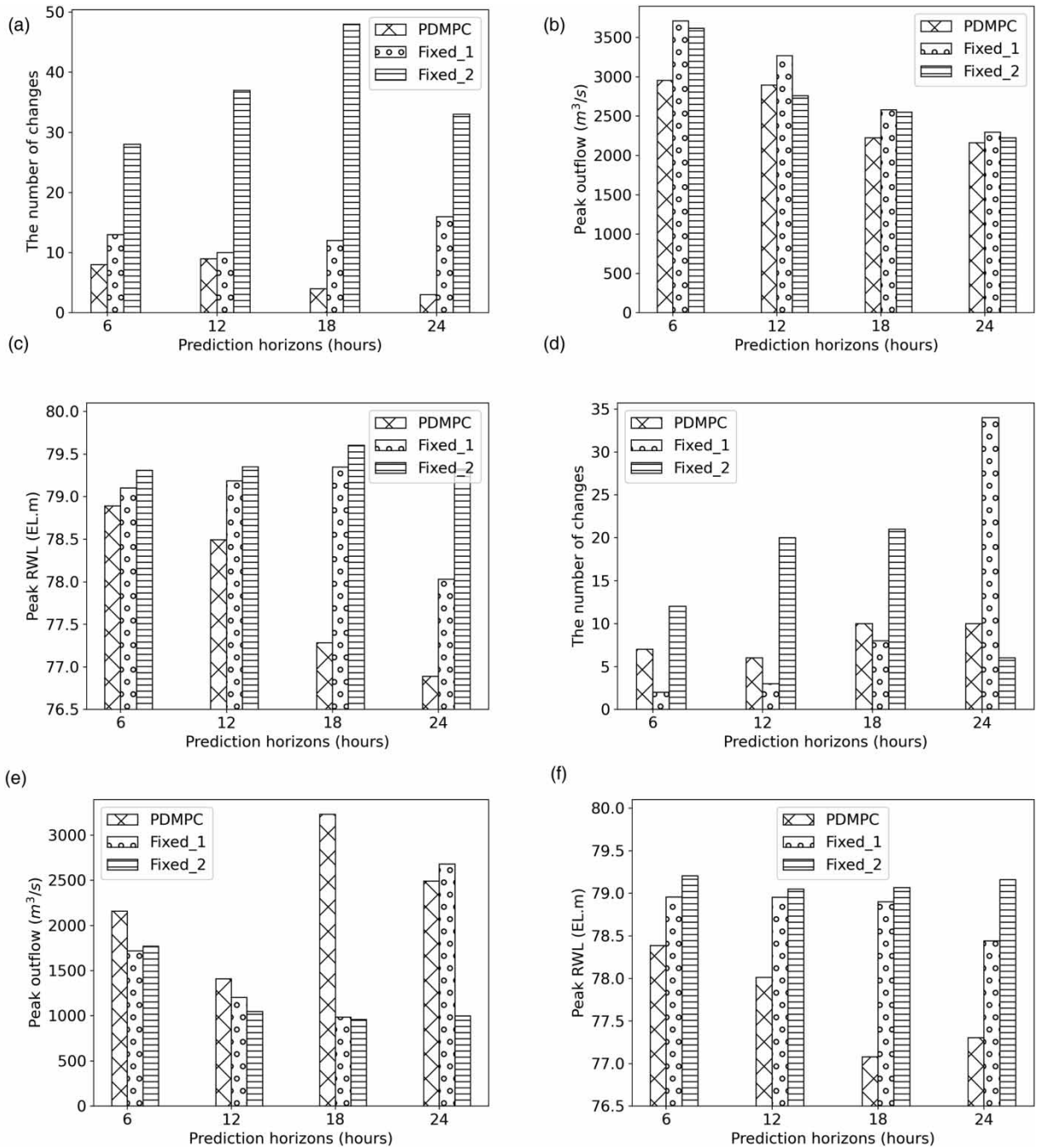

**Figure 4** | Comparison with the Fixed cases under uncertain inflow. (a) The number of changes between consecutive outflow schedules for Event 1. (b) Peak outflow for Event 1. (c) Peak RWL for Event 1. (d) The number of changes between consecutive outflow schedules for Event 3. (e) Peak outflow for Event 3. (f) Peak RWL for Event 3.

### 4.3. The best weights/parameters can vary with time

In Section 2.3, we discussed the complexity of trade-offs among objectives, which led us to assume that the operator preferences among them can be dynamic.

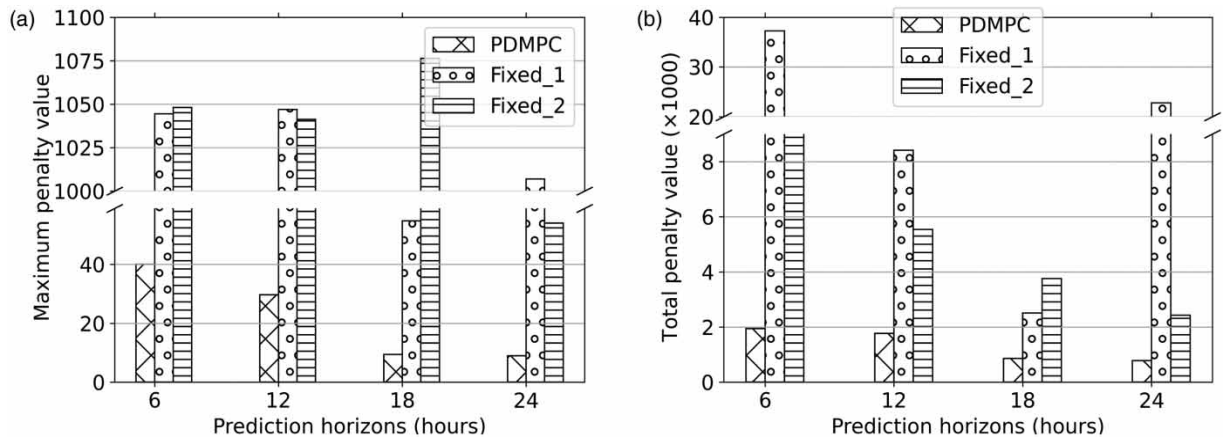

**Figure 5** | The penalty values from the evaluator for Event 1 under uncertain inflows. (a) The maximum penalty and (b) the total penalty (sum of penalty values).

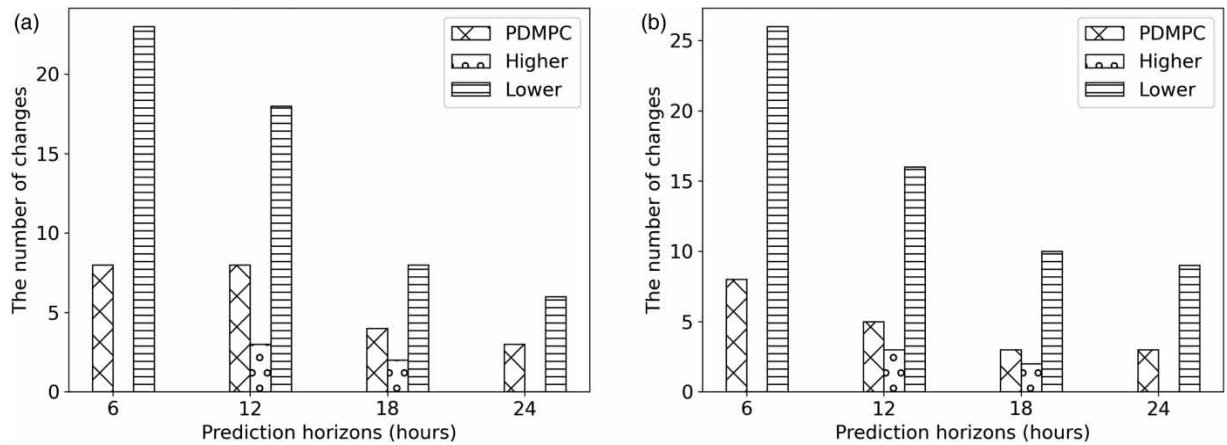

**Figure 6** | The number of increases with different evaluator settings for Event 1 (a) under uncertain inflow and (b) under certain inflow.

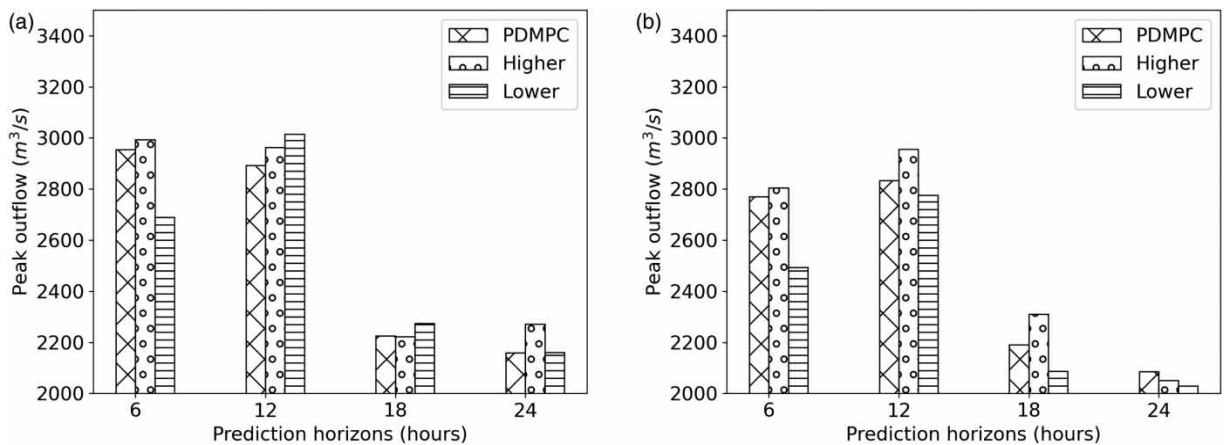

**Figure 7** | The peak outflow with different evaluator settings for Event 1 (a) under uncertain inflow and (b) under certain inflow.

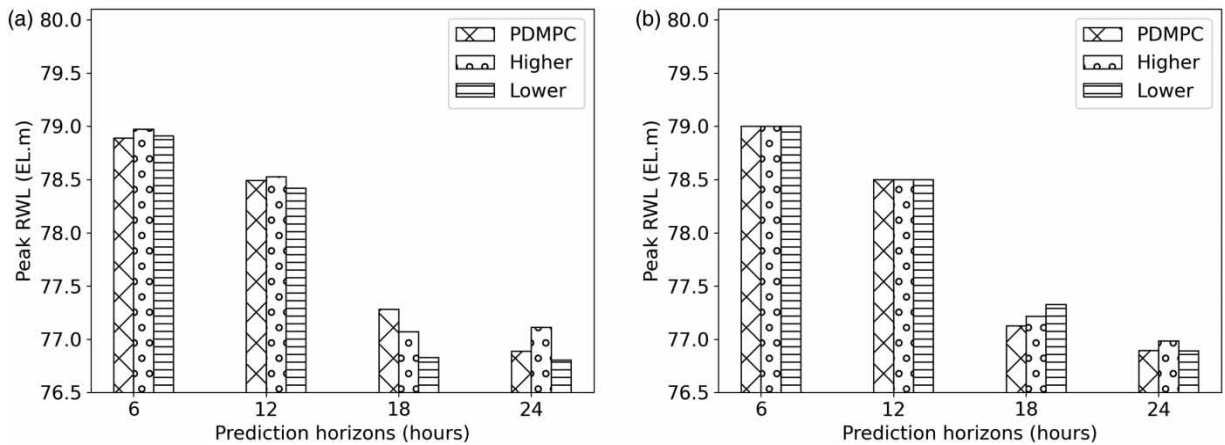

**Figure 8** | The peak RWL with different evaluator settings for Event 1 (a) under uncertain inflow and (b) under certain inflow.

Figure 10 illustrates the variation in penalty values as a weight only for the objective function related to RWL ($J5$) changes within our search range, while other objectives' weights are the same as the weights from PD-MPC, resulting in Figure 3. For example, when a weight value for the $J5$ is two, it denotes that $w_{5,1}$ is $2 \times \frac{20}{FWS \times F}$, $w_{5,2}$ is $2 \times \frac{40}{FWS \times F}$, and $w_{5,3}$ is $2 \times \frac{400}{FWS \times F}$ (see Table 3). Bright colors in the figure represent relatively low penalty values. The value of 99 in black color does not mean the penalty value is precisely 99, but the actual value is much higher. For simplicity, we convert all high penalty values into 99, which indicates undesirable conditions such as when the outflow exceeds the peak inflow up to the current time step or the RWL approaches FWL. Since the evaluator computes a nonlinear objective with nonlinear system simulations (see Figure 1), the same penalty values, which also mean the same control inputs, are observed in many cases for different weight sets. Despite that, it is evident that the best weight set changes at every time step. For example, in Figure 10(c), a low weight on $J5$ when $k < 90$ produces the lowest total objective value, but when $k = 90$, a high weight results in the lowest total objective value. This dynamic nature of optimal weight sets underscores the need for a flexible and adaptable approach in reservoir flood control. Additionally, it effectively conveys that the outputs from an MPC formulation with constant weights can, at best, be equal to those obtained through PD-MPC.

### 4.4. Discussion

It appears that the detailed design of practical objectives for reservoir flood control has not been thoroughly presented (Pianosi *et al.* 2020; Ritter *et al.* 2020), and it seems complicated to formulate practical objectives with only linear equations, as we described in Section 2.2. The receding optimal control of an MPC problem with numerous nonlinear objectives and constraints can become intractable to solve online (Allgower *et al.* 2004; Berberich *et al.* 2022). Given that preferences can be expressed as objective weights to represent their relative importance (Wang *et al.* 2017), the optimal preferences (weights) should adapt to varying hydrological conditions of the reservoir. However, it seems that the operators' preferences have not been adequately incorporated, and constant weights/parameters have been used to optimise reservoir flood control (Van Overloop 2006; Xu *et al.* 2011; Breckpot *et al.* 2013; Qi *et al.* 2017; Uysal *et al.* 2018b; Aydin *et al.* 2022).

In this study, we harness the advantages of solving linear MPC problems by parameterising the nonlinear and dynamic preferences around the different operating points. This allows us to optimise weights/parameters and control inputs simultaneously. Our approach is also in the spirit of multi-objective MPC methods for linear systems (Bemporad & de la Peña 2009) where a time-varying, state-dependent decision criterion can be taken into account using parametric optimisation. First, we present objectives for practical flood control in detail. Some of them have not been extensively covered in prior literature despite their significant importance in practice, such as minimising the changes in outflow schedules. Subsequently, we categorise these objectives into linear and nonlinear ones following the parameterisation of the operator's preference, i.e., the weights of objectives and parameters. We employ the genetic algorithm (GA) to optimise nonlinear objectives and constraints to derive the optimal weights/parameters of the linear MPC formulation at each time step. We refer to this framework as a parameterised dynamic model predictive control (PD-MPC).

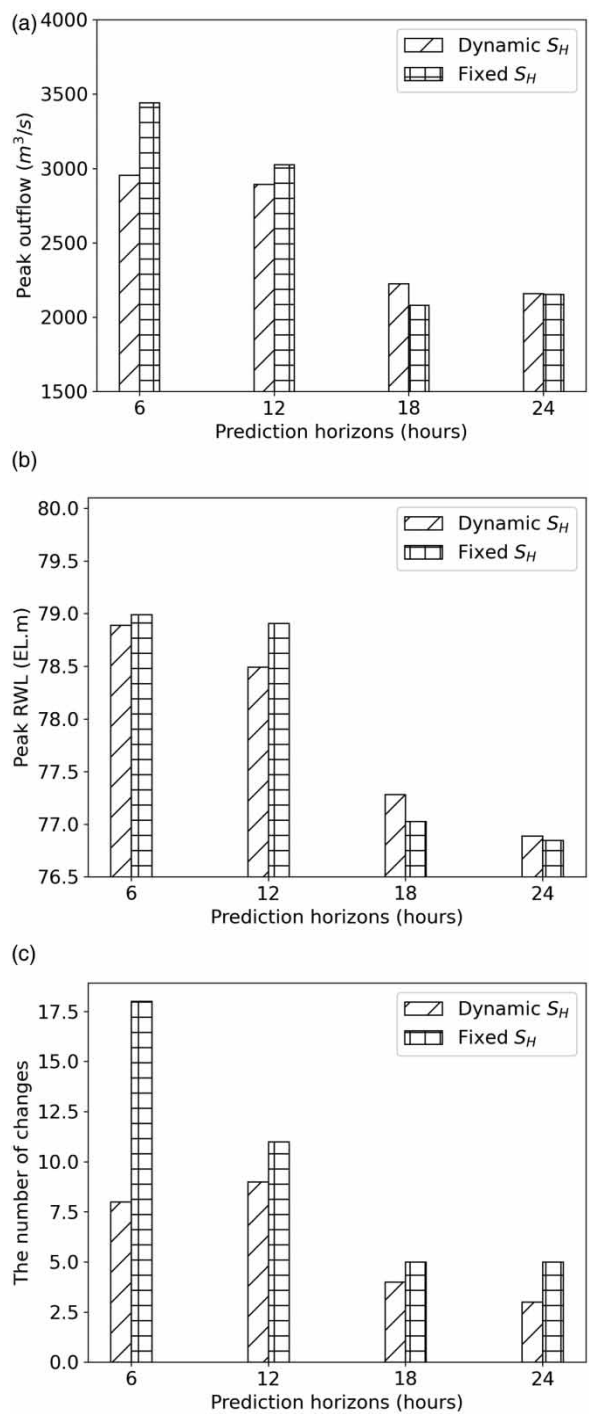

**Figure 9** | Comparison between dynamically changing $S_H$ and fixed $S_H$ under uncertain inflow. (a) Peak outflow. (b) Peak RWL. (c) The number of changes between consecutive outflow schedules.

Through our numerical experiment, we demonstrated that PD-MPC shows robustness to the inflow uncertainty. It is important to note that the PD-MPC framework does not directly handle inflow uncertainty. Instead, we rely on the robustness of the receding horizon MPC approach to address uncertainty and produce reliable results (De Nicolao *et al.* 1996; Schwenzer *et al.* 2021). Our results reveal that PD-MPC outperformed MPC formulations with fixed weights/parameters, even when these fixed weights/parameters were specifically designed for individual objectives. We showed that the weights/parameters in MPC formulations and the weights of objectives should also vary dynamically to adapt to changing conditions. PD-MPC then effectively adapted to the changing hydrological conditions and continuously updated the weights/parameters. This adaptability allowed it to make optimal decisions in real-time, resulting in better overall performance in terms of peak out-flows, reservoir water levels, and the number of changes in outflow schedules. This adaptability is the key factor in its superior performance and is also essential for generating optimal control inputs reflecting the dynamic characteristic of the operator preferences.

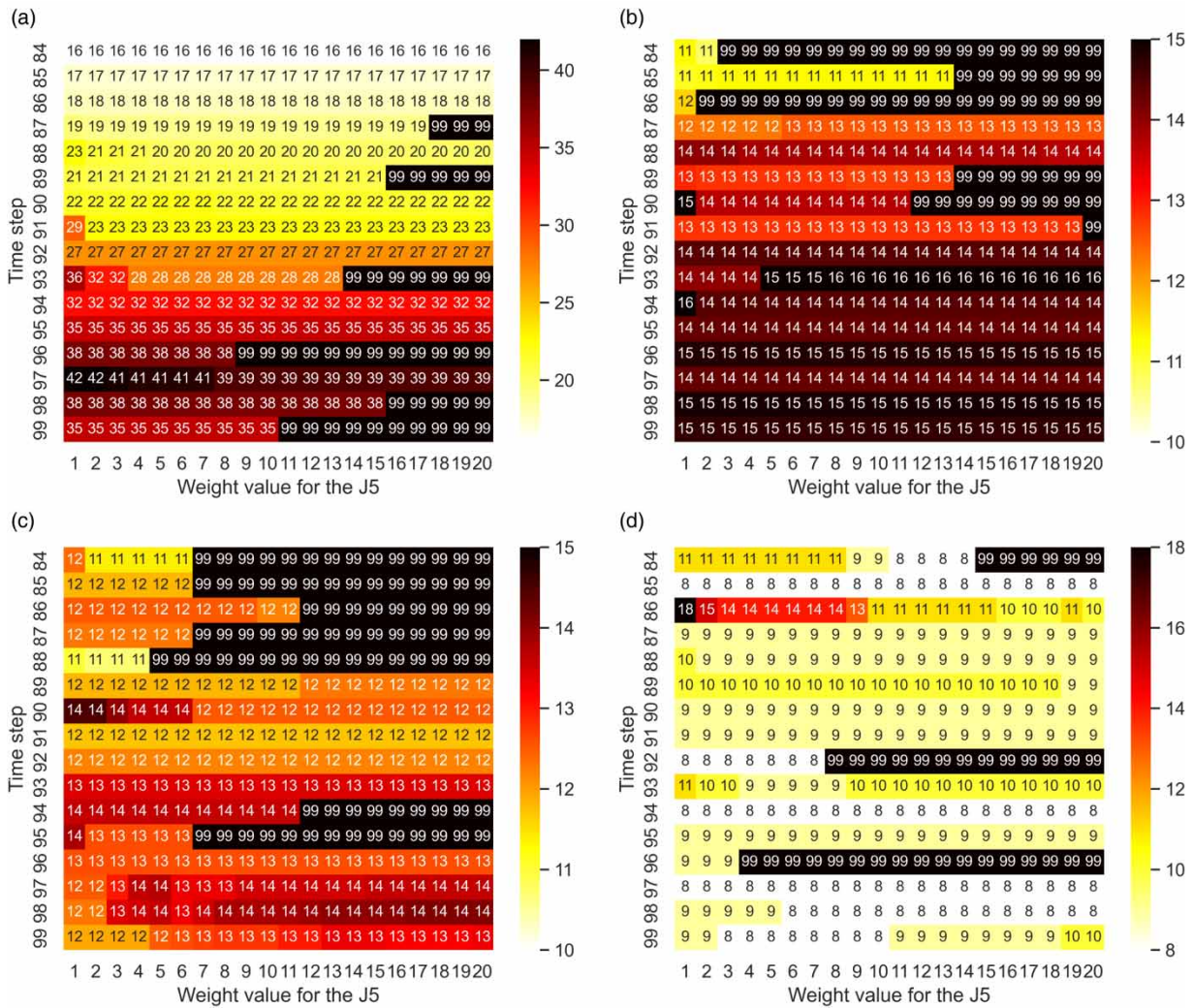

**Figure 10** | The colour represents the penalty value calculated at each time step from $k = 84$ to $k = 99$, with different weight values for the $J5$ while inflow increases steeply for Event 1. The prediction horizons are for: (a) 6 h. (b) 12 h. (c) 18 h. (d) 24 hours.

Nevertheless, this research has some limitations. First, the MPC formulation applied here does not consider the final states of the system. MPC with policy search algorithms (Song & Scaramuzza 2022) or a value function produced by a Reinforcement Learning (RL) model (Arroyo et al. 2022) can give a chance to find the approximation of the terminal cost such that our MPC framework can consider the whole period of a flood event. To the best of our knowledge, this approach has never been applied to a reservoir system for the purpose of flood control. The second limitation is related to uncertainty. We showed that PD-MPC can generate acceptable control inputs under uncertainty from the inherent feedback mechanism of a receding hor-izon implementation; however, it is advisable to explore PD-MPC with stochastic/robust MPC (Saltık et al. 2018) or Learning-based MPC (Hewing et al. 2020) to ensure robustness by considering uncertainty explicitly. The third limitation per-tains to our numerical experiments, which utilised three historical flood events. Although these events present significant operational challenges that demonstrate our method's practical utility, e.g., long event periods and two or three peaks, as well as a limitation in obtaining hydrograph of the 200-year flood and PMF, to establish broader applicability, numerical experiments under more extreme conditions are necessary. Finally, we did not consider the downstream impact of reservoir outflows as well as the upper reservoir for simplicity. This should be considered important in practical operations. The downstream impact can be considered using routing models or various heuristics suggested in many studies (Hsu & Wei 2007; Le Ngo *et al.* 2007; Peng *et al.* 2017). The entire reservoir system, including the upper reservoirs, needs to be explored, and we expect that PD-MPC could be applied to the joint optimisation problem, though this may be a topic for further study.

## 5. CONCLUSION

This study addresses the limitations of existing reservoir control approaches from a practical point of view, highlighting the current limitations of the employed optimisation approaches to take into account specific operators' preferences, which may change over time. We assume dynamic preferences by operators in a multi-objective setting and show the dynamic characteristics of weights/parameters. We then propose a PD-MPC framework as a parameterised linear MPC with dynamic optimisation of weights/parameters via a model-based learning concept. We applied this methodology to the Daecheong reservoir, verified the dynamic-preference assumptions, and tested this framework's effectiveness. This study lays the foundation for developing more adaptive decision-making frameworks in reservoir flood control and other related fields, being closer to the actual set of preferences in reservoir management and hence having the level of adoption by practitioners.

Despite the mentioned limitations of this study, we think it allows for making one step towards wider adoption of optimisation approaches to real-time reservoir flood control. The presented methodology is, of course, not aimed at replacing manual operation but rather gives instruments for reducing operators' stress in critical situations and ultimately enhancing their ability to make better decisions.

## ACKNOWLEDGEMENTS

We express our gratitude to Korea Water Resources Public Corporation (K-water) for their sponsorship of the first author and for sharing the data and information related to the Daecheong reservoir. The hydrological and operational data of the Daecheong reservoir is accessible to the public on K-water's website (http://kwater.or.kr). We also thank the ICT Cooperative of Dutch Education and Research Institutions (SURF) for allowing us to utilise the Dutch national e-infrastructure with the support of the SURF Cooperative using grant no. EINF-6342.

## DATA AVAILABILITY STATEMENT

All relevant data are available from an online repositoryor repositories (please ensure the DOI/URL has been provided as a submssion item).

## CONFLICT OF INTEREST

The authors declare there is no conflict.

corrected Proof